\documentstyle[preprint,aps]{revtex}
\begin{document}
\draft

\title{Neutrino spin-flips in curved space-time}

\author{M. Br\"uggen} \address{Institute of Astronomy, University of
Cambridge, Madingley Road, Cambridge CB3 0HA, UK} 

\maketitle

\begin{abstract}
The general relativistic effects on spin-flavor oscillations above the
core of type II supernovae are investigated. The evolution equation is
derived and the relative magnitudes of the terms in the Hamiltonian,
which arise from the weak, electromagnetic and gravitational
interaction, are compared. The effects on the resonance position and
the adiabaticity are studied. Explicit calculations are presented for
non-rotating and slowly rotating stars.
\end{abstract}

\pacs{14.60.Pq, 95.30.Sf, 26.30.+k, 97.60.Bw}

\section{Introduction}

If the neutrino possesses a magnetic moment, it has been shown that
its spin can be flipped while passing through a magnetic field
\cite{LiMa,Vo,Akh2,SV,VVO}. This can lead to spin-precession and
spin-flavor oscillations.  In the standard model of particle physics
the neutrino magnetic moment is zero. However non-standard neutrino
properties such as mass and magnetic moment attract enormous attention
since they are not only natural by-products of many grand unified
theories; but they can also lead to neutrino oscillations, which
provide a very popular and elegant solution to the long-standing solar
neutrino problem \cite{haxton}.

Terrestrial experiments constrain the magnetic moment of the neutrino
to values $<10^{-10}\mu_B$ \cite{PDG,Vid,VoEn}, where $\mu_B$ is the
Bohr magneton.  Astrophysical bounds are more stringent constraining
$\mu < 2\times 10^{-12}\mu_B$ \cite{No,Pe,LaCo,RaWe,Ra}. The latter
are, however, somewhat model-dependent.

Thus for spin-flips to be discernible, very strong magnetic fields are
required. One of the places where magnetic fields of such strengths
are likely to exist is above the proto-neutron star of a type II
supernova (SN). Observations of magnetic fields on the surface of
white dwarfs suggest the existence of magnetic fields of up to
$10^{10}$ G at the surface of the iron core \cite{Cha}; energetic
considerations suggest even higher fields \cite{ToSa}.

In type II SNe neutrinos play a crucial role since they are
believed to drive the supernova explosion \cite{BW,Bu}. Neutrino
oscillations could therefore affect the explosion mechanism
directly. Also neutrinos from a nearby SN explosion, SN 1987 A, have
been detected on Earth \cite{Hi}, and the measured neutrino spectrum
has been used to test various oscillation scenarios \cite{Lo,Sm}.
Neutrino spin-flips have also been invoked to explain the high pulsar
kick velocities, as the interaction with a magnetic field can lead to
an asymmetric neutrino emission \cite{Akh}. 

It is known for some time that gyroscopes precess in strong
gravitational fields \cite{Th,BrCoh}. Here I will investigate under
which conditions gravitational effects can influence the neutrino spin
flip in a SN.

SN explosions occur at the end of the life of massive stars and leave
a neutron star or a black hole behind. Therefore it is of importance to
study neutrino oscillations in curved space-time and calculate the
magnitude of general relativistic effects. Cardall \& Fuller
\cite{CF} have calculated the gravitational effect on neutrinos in
vacuum and matter. They derived MSW (Mikheyev-Smirnov-Wolfenstein)
oscillation formulae in curved space-time and presented expressions
for radial and azimuthal propagation in a Schwarzschild
geometry. Piriz {\it et al.} \cite{PRW} derived the resonance
conditions for neutrino conversions with regard to neutrino
oscillations in the vicinity of active galactic nuclei. They present
explicit calculations for arbitrary orbits in a Kerr-geometry.

Here I will briefly rederive the evolution equation for massive Dirac
neutrinos interacting with a magnetic field in matter. In Sec. 3 and 4
I will study how the neutrino conversion is affected by the curvature
of space-time near a static and a slowly rotating proto-neutron star.

Throughout this paper I shall set $G=\hbar=c=1$.

\section{The evolution equation in curved space-time}

Gravitational effects on the spin arise through the ``spin
connection'' $\Gamma_{\mu}$ appearing in the Dirac equation. This
effect has been studied by Cardall \& Fuller \cite{CF} and somewhat
more formally by Piriz {\it et al.} \cite{PRW} (also see
\cite{ahl,gross}). In the following I will follow the treatment of Ref.
\cite{CF}. The Dirac equation in curved space-time (ignoring magnetic
fields and matter effects for the moment, and restricting myself to
two neutrino species) for massive neutrinos interacting minimally with
the gravitational field can be written as follows \cite{weinberg2}:
\begin{equation}
\left[ \gamma^{a} e^{\mu}_a (\partial_{\mu} + \Gamma_{\mu}) 
	+ M 
	\right]
	\psi = 0.  \label{1}
\end{equation}
Here $\psi$ is a column vector of spinors of different mass
eigenstates, the $\gamma^{a}$'s represent the Dirac matrices and $M$
is the mass matrix, $M^2={\rm diag}[m_1^2,m_2^2]$, where $m_1$ and
$m_2$ are the mass eigenvalues of the two neutrino species. I follow
the convention that greek indices refer to general curvilinear
coordinates, while the latin indices $a,b,c,d$ refer to locally
inertial (Minkowski) coordinates. The tetrads $e^{\mu}_a$ connect
these sets of coordinates. $g_{\mu\nu}$ will denote the metric of
curved space-time and $\eta_{ab}$ the metric in flat space, where I
will use the convention that $\eta_{ab}={\rm diag}[-1,1,1,1]$. The
tetrads then satisfy
\begin{equation}
g_{\mu\nu}=\eta_{ab} e_{\mu}^a e_{\nu}^b.\label{2}
\end{equation}
The explicit expression for $\Gamma_{\mu}$ is
\begin{equation}
\Gamma_{\mu} = {1 \over 8}[\gamma^b, \gamma^c] e^{\nu}_b
	e_{c \nu;\mu}.
\end{equation}
Here the semicolon denotes a covariant derivative and a comma an
ordinary derivative.  It can be shown that
\begin{equation}
\gamma^a [\gamma^b, \gamma^c] = 2 \eta^{ab} \gamma^c -
	2 \eta^{ac} \gamma^b - 2i \epsilon^{dabc} \gamma_5 
	\gamma_d, \label{gammas}
\end{equation}
where $\epsilon^{abcd}$ is the (flat space) totally antisymmetric
tensor, with $\epsilon^{0123}= +1$. With Eq. (\ref{gammas}), the
non-vanishing contribution from the spin connection is
\begin{equation}
\gamma^a e_a^{\mu} \Gamma_{\mu} = \gamma^a e_a^{\mu}
	\left\{  i
	A_{G\mu} \left[-{\gamma_5 \over 2}\right] \right\},
	\label{3}
\end{equation}
where 
\begin{equation}
A^{\mu}_G \equiv {1 \over 4}  e^{\mu}_a \epsilon^{abcd}
	(e_{b\nu,\sigma} - e_{b\sigma,\nu}) e^{\nu}_c
	e^{\sigma}_d.\label{4}
\end{equation}
The expression in Eq. (\ref{3}) treats left- and right-handed
states differently. In order to group it with terms arising
from matter effects, we can without physical consequence
add a term proportional to the identity to obtain
\begin{equation}
\gamma^a e_a^{\mu} \Gamma_{\mu} = \gamma^a e_a^{\mu}
	(i A_{G\mu} {\cal P}_L),\label{5}
\end{equation}
where ${\cal P}_L$ is the left-handed projection operator ${\cal
P}_L=(1-\gamma^5)/2$. As pointed out by Cardall \& Fuller \cite{CF},
$A_{G\mu}$ is diagonal in spin space and cannot produce spin flips on
its own. However it can change the resonance as well as the
adiabaticity conditions in the presence of other off-diagonal terms
such as those arising from an interaction with a magnetic field.

A further complication arises in curved space-time: The neutrino
trajectories are no longer necessarily straight lines as in flat
space-time. They can be written as curves of the affine parameter
$\lambda$. The tangent vectors to the null worldline,
$p^{\mu}=dx^{\mu}/d\lambda$, can be found by solving the geodesic or
Hamilton-Jacobi equation (see, e.g.,\cite{LL}). As suggested in Ref.
\cite{CF} the four-momentum $P_{\mu}$ is then constructed by assuming
that the neutrinos are energy eigenstates and that the three momentum
is collinear with $(p_1,p_2,p_3)$.\\

The mass matrix $M$ is related to the mass matrix in flavor space
$M_f$ through a rotation by the vacuum mixing angle $\theta_v$ as
follows:

\begin{equation}
M_f^2 = U \pmatrix{m_1^2 &0 \cr 0& m_2^2} U^{\dagger},
\end{equation}\label{6}
with
\begin{equation}
U = \pmatrix{\cos \theta_v & \sin \theta_v \cr -\sin \theta_v
        & \cos \theta_v}.\label{7}
\end{equation}

The neutrinos interact with matter through a weak interaction current
of the form: $A_w^{\mu}{\cal P}_L$, where in the chiral basis
($\nu_{eL}$,$\nu_{eR}$,$\nu_{\mu L}$,$\nu_{\mu R}$): $A_w^{\mu}={\rm
diag}[V_c+V_n,V_n,0,0]$ (Here the $\nu_{\mu}$s can, of course, be
replaced by $\nu_{\tau}$s.). The matter potentials due to the neutral
and charged current in neutral, unpolarized matter are given by
$V_n=-\sqrt{2}G_F n_n $ and $V_c=2\sqrt{2}G_F n_e $,
respectively, where $G_F$ is Fermi's constant and $E_l$ the energy as
measured in a local inertial frame. $n_e$ and $n_n$ are the electron
and neutron density in the rest frame of the background matter.\\

The interaction of the neutrino with an external magnetic field is
described by a term of the form
\begin{equation}
\hat{\mu}\sigma^{ab}F_{ab}\psi ,
\end{equation}
where $\hat{\mu}$ is the magnetic moment matrix, $F_{ab}$ the
electromagnetic field tensor and
$\sigma^{ab}=\frac{1}{4}[\gamma^a,\gamma^b]$.

Only considering conversions between two neutrino flavors we can
write the evolution equation as follows:

\begin{equation}
i\frac{d}{d\lambda}\left ( 
\begin{array}{c} \nu_{eL} \\ \nu_{\mu L} \\ \nu_{eR} \\ \nu_{\mu R}
\end{array}
\right ) = \hat{H} \left ( 
\begin{array}{c} \nu_{eL} \\ \nu_{\mu L} \\ \nu_{eR} \\ \nu_{\mu R}
\end{array}
\right ),\label{8}
\end{equation}
where $\hat{H}$ is the Hamiltonian in the chiral basis. Including the
weak, electromagnetic and gravitational interaction the Hamiltonian
can be written as:

\begin{equation}
\hat{H} = \left (
\begin{array}{cccc} V_c+V_n+P_{\mu}A^{\mu}_G/E_l-\frac{\delta m^2}
{4E_l}\cos 2\theta_v & \frac{\delta m^2}{4E_l}\sin 2\theta_v &
\mu_{ee}B & \mu_{e\mu}B\\

\frac{\delta m^2}{4E_l}\sin 2\theta_v &
V_n+P_{\mu}A^{\mu}_G/E_l+\frac{\delta m^2}{4E_l}\cos 2\theta_v &
\mu_{\mu e}B & \mu_{\mu\mu}B \\

\mu_{ee}B & \mu_{e\mu}B & -\frac{\delta m^2}{4E_l}\cos 2\theta_v &
\frac{\delta m^2}{4E_l}\sin 2\theta_v\\ \mu_{\mu e}B & \mu_{\mu\mu}B &
\frac{\delta m^2}{4E_l}\sin 2\theta_v & \frac{\delta m^2}{4E_l}\cos
2\theta_v
\end{array}\right ),\label{9}
\end{equation}
where $B$ is the component of the magnetic field which is
perpendicular to the neutrino trajectory and $\delta m^2$ the
difference of the squares of the neutrino mass eigenvalues in matter:
$\delta m^2=m_2^2-m_1^2$. Here it was assumed that the right-handed
components are sterile and do not interact with matter. Resonance
occurs whenever the diagonal elements cross. From the expression for
$\hat{H}$ we can immediately read off the resonance conditions for
spin-flips and MSW-conversions:\\ For $\nu_{eL}\leftrightarrow\nu_{\mu
L}$:

\begin{equation}
V_c-\frac{\delta m^2}{2E_l}\cos 2\theta_v=0,\label{10}
\end{equation}
for $\nu_{eL}\leftrightarrow\nu_{eR}$:

\begin{equation}
V_c+V_n+P_{\mu}A^{\mu}_G/E_l=0,\label{11}
\end{equation}
for $\nu_{eL}\leftrightarrow\nu_{\mu R}$:

\begin{equation}
V_c+V_n+P_{\mu}A^{\mu}_G/E_l-\frac{\delta m^2}{2E_l}\cos
2\theta_v=0,\label{12}
\end{equation}
for $\nu_{\mu L}\leftrightarrow\nu_{eR}$:

\begin{equation}
V_n+P_{\mu}A^{\mu}_G/E_l+\frac{\delta m^2}{2E_l}\cos
2\theta_v=0,\label{13}
\end{equation}
and for $\nu_{\mu L}\leftrightarrow\nu_{\mu R}$:

\begin{equation}
V_n+P_{\mu}A^{\mu}_G/E_l=0 .\label{14}
\end{equation}
These equations are similar to those derived in Ref. \cite{PRW}. In
the active-active MSW transition, Eq. (\ref{10}), the matter potential
is due to the charged current only and proportional to $Y_e$. [$Y_e$
is the electron fraction: $Y_e=n_e/(n_n+n_e)$.]  However for the
transitions corresponding to Eqs. (\ref{11}) and (\ref{12}) the
potential is due to the neutral as well as the charged current and
thus proportional to $(1-2Y_e)$. The latter potential is strongly
suppressed in the isotopically neutral region (number of
protons=number of neutrons) \cite{At}.

In the following sections I will calculate the `gravitational
current', $P_{\mu}A^{\mu}_G$, for a static as well as for a rotating
star.

\section{Relativistic effects near static star}

The geometry in a spherically symmetric, static spacetime can be
globally represented by the Schwarzschild coordinate system
$\{x^{\mu}\}\Rightarrow (t,r,\theta,\phi)$. MSW transitions in a
Schwarzschild metric have already been discussed in Ref. \cite{CF} and
their findings can easily be extended to spin-flip conversions.  The
Schwarzschild line element, which serves to define these coordinates,
is
\begin{eqnarray}
ds^2 &=& g_{\mu\nu}dx^{\mu}dx^{\nu} \nonumber \\ 
	&=& -e^{2\Phi(r)} dt^2 + e^{2\Lambda(r)} dr^2 \nonumber \\
	& &+ r^2 d\theta^2 + r^2 \sin ^2\theta d\phi ^2.  \label{15}
\end{eqnarray}
Using the tetrads
\begin{equation}
e^{\mu}_a = {\rm diag}\left[ e^{-\Phi (r)},
	e^{-\Lambda (r)}, {1\over r}, 
	{1 \over r \sin \theta} \right],	\label{16}
\end{equation}
it is easy to verify that $A_G^{\mu}=0$, as pointed out by
\cite{CF}. In Schwarzschild metric it is common to write $e^{2\Phi
(r)}=1-r_s/r$ and $e^{2\Lambda (r)}=1/(1-r_s/r)$, where the
Schwarzschild radius $r_s=2{\cal M}$ and ${\cal M}$ is the central
gravitating mass. The energy-like component of the four-momentum is
the energy and a conserved quantity: $P_0=-E$. The energy measured by
a locally inertial observer is $E_l=Ee^{-\Phi (r)}$; i.e., the energy
of the neutrino suffers a gravitational redshift by $1/\sqrt{1-r_s/r}$
\cite{Fu}.  Thus the resonance conditions are merely altered by
setting $E_l$ equal to $Ee^{-\Phi (r)}$, and $A_G^{\mu}=0$.

For a particular transition one can define a helicity mixing angle,
$\theta_m$, which diagonalises the corresponding submatrix in the
Hamiltonian (see, e.g., \cite{haxton}). For instance, for the
transition $\nu_{eL}\leftrightarrow\nu_{\mu R}$, the mixing angle is
defined by:

\begin{equation}
\tan 2\theta_m = \frac{2\mu_{e\mu}B}{e^{\Phi} \delta m^2\cos
2\theta_v /2E - (V_c+V_n)}.\label{17}
\end{equation}
The adiabaticity parameter corresponding to a particular transition,
$\gamma$, quantifies the magnitude of the off-diagonal elements
compared to the diagonal ones of $\hat{H}$ in the basis of the
instantaneous eigenstates. At the resonance point it must be greater
than 1 for adiabatic propagation. $\gamma$ is modified by relativistic
effects, and for the case $\nu_{eL}\leftrightarrow\nu_{\mu R}$, can be
written as:

\begin{equation}
\gamma (r_{res})=\frac{8e^{-\Phi(r_{res})}E\mu^2_{e\mu}B^2}{\delta
m^2\cos 2\theta_v e^{-\Lambda(r_{res})}|d\ln (V_c+V_n)/dr|},\label{18}
\end{equation}
where the logarithmic derivative is evaluated at the resonance
position.  Thus the adiabaticity of the evolving neutrino is affected
by the spatial dependence of the metric, which appears for example in
the determination of the local energy. The conversion probability can
then be expressed as \cite{haxton}
\begin{equation}
P_{\nu_{eL}\rightarrow\nu_{\mu R}}=\frac{1}{2}-\left
(\frac{1}{2}-P_{hop}\right )\cos 2\theta_v\cos 2\theta_m ,\label{19}
\end{equation}
where the non-adiabatic hopping probability is given by
$P_{hop}=\exp\left[-{\pi \over 2} \gamma (r_{\rm res})\right]$.

The conversion probabilities corresponding to the other transitions of
Eqs. (\ref{10}-\ref{14}) can be found accordingly.

\section{Relativistic effects near rotating star}

Rotation of the collapsing star will cause a dragging of the inertial
frames and will result in off-diagonal terms in the metric
$g_{\mu\nu}$. These off-diagonal terms can now flip the spin of the
neutrinos by transferring the angular momentum of the star to the
neutrino spin. Using perturbation theory one can write down a metric
to first order in the angular velocity of rotation $\Omega$ of the
central mass (e.g., see \cite{Har,MTW}):
\begin{eqnarray}
ds^2 &=& -e^{2\Phi(r)} dt^2 + e^{2\Lambda(r)} dr^2 \nonumber \\ & &+
	r^2(d\theta^2 + \sin^2\theta d\phi ^2) -2 \omega
	r^2\sin^2\theta d\phi dt, \label{20}
\end{eqnarray}
where $\omega$ is the local precession frequency and can be written as
$\omega=2J/r^3$ \cite{Har}, where $J$ is identified with the angular
momentum of the central gravitating mass. Here it is assumed that the
angular momentum vector of the star, ${\bf J}$, is aligned with the polar
axis. The metric of Eq. (\ref{20}) is valid as long as the rotation is
slow, i.e., for $\Omega a<<1$, where $a$ is the radius of the central
mass.

In this metric purely radial motion is only possible along the axis
of rotation as expected from symmetry considerations. From these same
considerations it is also evident that motion in a plane is only
possible in the equatorial plane.

However, here I will confine myself to nearly radial trajectories and
work in the weak-field limit so that the curvature of the neutrino
path can be neglected. Thus one finds from Eqs. (\ref{4}) and
(\ref{20}):

\begin{equation}
P_{\mu}A^{\mu}_G/E_l\approx e^{-\Lambda
(r)}Jr^{-3}\cos\theta ,\label{21}
\end{equation}
for relativistic neutrinos moving radially outwards (i.e. $P_1\approx
P_0$).  It can be noted that for propagation in the equatorial plane
the `gravitational current' is still zero. But at any other
polar angle there will now be an additional term in the Hamiltonian
which will shift the resonance position and modify the adiabaticity
conditions. In Eqs. (\ref{17}) and (\ref{18}) one can then simply
replace $V_c+V_n$ by $V_c+V_n+P_{\mu}A_G^{\mu}/E_l$.  It is
interesting to note that $P_{\mu}A_G^{\mu}$ changes sign at the
equator and could thus lead to an asymmetry in the neutrino
emission. It is evident from Eq. (\ref{21}) that the `gravitational
current' moves the resonance point closer to the center of the SN in
the upper hemisphere and farther away in the lower hemisphere. The
magnitude of this effect will be discussed in the following section.

\section{Discussion}

Simple angular momentum arguments suggest high angular velocities for
neutron stars, which have also been verified by observations. During
collapse to a neutron star the angular velocity of a typical star will
increase from about $\Omega_0\approx 10^{-6} {\rm s}^{-1}$ to
$\Omega\approx 10^{4} {\rm s}^{-1}$. Fig. 1 shows the absolute
magnitudes of the various components in the Hamiltonian given by
Eq. (\ref{9}) for a neutrino propagating in a typical SN
environment. I consider here the case when the proto-neutron star is
rotating, as $A_G=0$ otherwise. The neutrino is assumed to move
radially along the polar axis. The matter potential was
calculated using the progenitor model for a 15 $M_{\odot}$, low
metallicity star as calculated in Ref. \cite{ToSa} on the basis of the
models of Ref. \cite{WW}. It can be seen that in the region between
0.02 $R_{\odot}$ and 0.3 $R_{\odot}$ the matter potential is strongly
suppressed due to the high abundance of isotopic elements such as
Oxygen, Carbon and Helium. 

The magnetic field was assumed to be a dipole field which is aligned
with the rotation axis and is given by:

\begin{equation}
B=B_0\left ( \frac{r_0}{r}\right )^3 \sin\theta,
\end{equation}
where $B_0$ is the magnitude of the magnetic field at the surface of
the iron core at a radius $r_0$. The value of the neutrino magnetic
moment was taken to be $\mu=10^{-12}\mu_B$ and $B_0$ was set to
$10^{10}$ G. The `gravitational current' was calculated assuming that
the iron core has collapsed to form a proto-neutron star of mass $M=2
M_{\odot}$ and a radius of $a=10$ km. Its angular velocity was taken
to be $\Omega=10^6 {\rm s}^{-1}$. It is seen that the `gravitational
current' comes relatively close to the matter potential in the
isotopically neutral region of the SN envelope. Also shown in Fig. 1
is the absolute difference of the diagonal elements of the Hamiltonian
for the transition $\nu_{eL}\leftrightarrow\nu_{\mu R}$, i.e. the
right-hand side of Eq. (\ref{11}). Resonance occurs whenever the
difference in the diagonal elements goes through zero. It is evident
that in order for the `gravitational current' to affect the neutrino
spin-flip, the resonant transition has to take place in the
isotopically neutral region. For a choice of $\delta m^2=10^{-10}$
eV$^2$ and $\cos\theta_v\approx 1$ (large mixing angle) the resonance
occurs at about $r=0.15 R_{\odot}$.  

For magnetic fields as high as indicated in Fig. 1 the off-diagonal
term, $\mu B$, dominates the potential by several orders of magnitude
such that in the isotopically neutral region spin-flavor precession of
almost maximum depth occurs. Strong precession occurs whenever $\mu B$
is much greater than $\Delta H$ and the precession amplitude is given
by \cite{ToSa}:
\begin{equation}
A_p=\frac{(2\mu B)^2}{(2\mu B)^2+\Delta H^2}
\end{equation}
In that case the gravitational current has a negligible effect on the
conversion probability. However for weaker - and possibly more
realistic -  magnetic fields ($B_0<10^7$ G) $\mu B$ becomes comparable
to $\Delta H$, suppressing the spin-flavor precession. However $\mu B$
is still greater than $\Delta H$ at the resonance layer, where
resonant conversion occurs as long as the adiabaticity condition is
satisfied.

It is readily seen that for our example the `gravitational current' is
still by about one order of magnitude smaller than the matter
potential and will thus not significantly affect the resonance
position or the adiabaticity in the envelope of the SN (even for
$B_0<10^7$ G) .

However if the mass of the proto-neutron star or its rotation rate is
increased, or the matter potential in the envelope is suppressed by
about one order of magnitude (i.e., by reducing the abundance of high
Z metals), the presence of the gravitational current will start to
shift the resonance position and affect the transition probability
noticably.  

The conditions for which the gravitational contribution
becomes important are neither contrived nor unrealistic. These
calculations are only intended to yield crude orders of magnitude
estimates for the different terms in the evolution equation and more
exact numbers would require a more careful modelling of the collapsing
star. For instance, I have neglected the infall velocity of the matter
during the collapse and only considered radial neutrino trajectories
in the weak-field limit. However these simple estimates presented here
show that the terms arising from the space-time curvature can
potentially become significant for spin-flip transitions in the SN
envelope even though they are probably not important in most realistic
cases. The extent to which they affect the neutrino transition depends
on the detailed composition of the SN, as well as the mass and
rotation rate of the proto-neutron star.

\acknowledgements

I am grateful to Dr. T. Totani for providing me with the data for the
progenitor models.

\begin{figure}
\caption{Magnitudes of the various components in the Hamiltonian of
(\ref{9}) as a function of distance from the center of a 15
$M_{\odot}$ type II SN. The energies are measured in eV and the radius
is measured in units of solar radii. The lines represent the matter
potential (solid), the gravitational term (short-dash), the magnetic
term (dotted) and the absolute difference of the diagonal elements in
the Hamiltonian (long-dash). }
\label{fig1}
\end{figure}

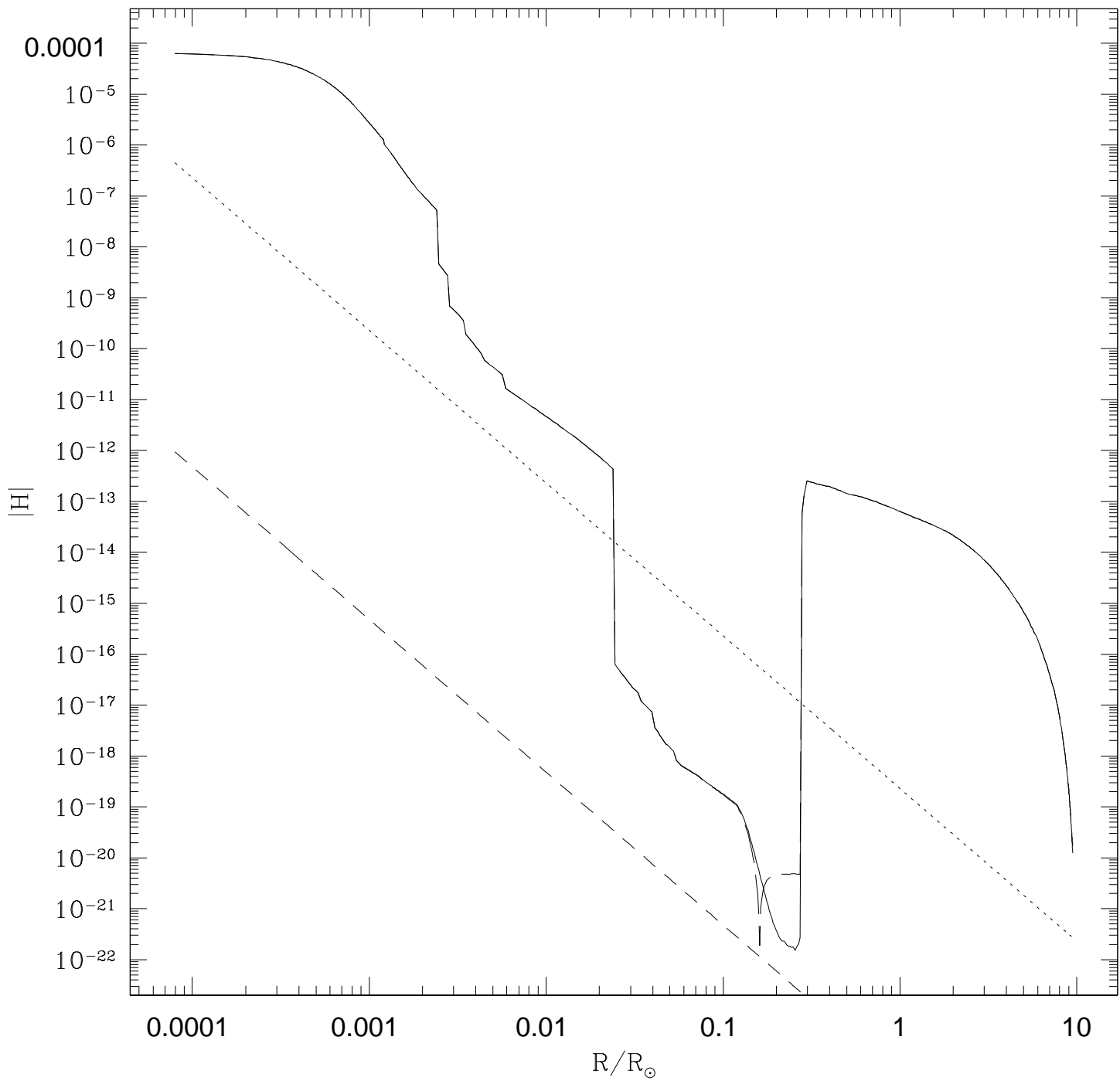
\end{document}